\documentclass[a4paper,12pt]{article}
\usepackage{latexsym}
\usepackage{amsmath}
\usepackage{accents}
\usepackage{bm}
\usepackage[T2A]{fontenc}
\usepackage[english]{babel}
\usepackage{graphicx}
\usepackage[usenames]{color}
\usepackage{colortbl}
\usepackage[font=footnotesize,figurewithin=none]{caption}
\usepackage{floatrow}
\usepackage[toc,page,title]{appendix}
\usepackage{CJK}
\usepackage{multirow}
\usepackage{subfigure}

\begin{document}

\def\Re{\mathop{\rm Re}\nolimits}
\def\Im{\mathop{\rm Im}\nolimits}

\centerline {\LARGE{Measuring entanglement of a rank-2 mixed state}}
\medskip
\centerline {\LARGE{prepared on a quantum computer}}
\medskip
\centerline {A.R.~Kuzmak$^1$, V.M.~Tkachuk$^2$}
\centerline {\small \it E-Mail: andrijkuzmak@gmail.com$^1$, voltkachuk@gmail.com$^2$}
\medskip
\centerline {\small \it Department for Theoretical Physics, Ivan Franko National University of Lviv,}
\medskip
\centerline {\small \it 12 Drahomanov St., Lviv, UA-79005, Ukraine}
\medskip

\date{\today}
\begin{abstract}
We study the entanglement between a certain qubit and the remaining system in rank-2 mixed states
prepared on the quantum computer. The protocol, which we propose for this purpose, is based on the relation
of geometric measure of entanglement with correlations between qubits. As a special case, we consider a two-qubit rank-2
mixed state and find the relation of concurrence with the geometric measure of entanglement. On the ibmq-melbourne
quantum computer we measure the geometric measure of entanglement in the cases of 2- and 4-qubit mixed quantum
states which consist of Schr\"odinger cat states. We study the dependence of the value of entanglement on the parameter
which defines the weight of pure states. Finally, we determine the concurrence of 2-qubit mixed state.
\end{abstract}

\section{Introduction \label{intro}}

Quantum computers are devices which allow implementing different quantum information schemes.
Unlike classical computers, quantum computers can effectively and rapidly implement a simulation of many-body quantum
systems \cite{feynman1982,lloyd1996,buluta2009,preskill2012}. Quantum computers can be used for studying different problems
related to the behavior of quantum systems in condensed-matter physics \cite{krishnan2019}, high-energy physics \cite{martinez2016},
atomic physics \cite{dumitrescu2018}, and quantum chemistry \cite{sugisaki2019}. It is possible because physical systems used for the implementation of quantum devices
are many-body controlled quantum systems such as spins of atoms \cite{loss1998,kane1998},
trapped atoms \cite{molmer1999,porras2004}, ultracold atoms \cite{duan2003,kuklov2003,gross2017}
or superconducting circuits \cite{wei2005,mooij1999,makhlin2001,majer2007}. An important property of these systems that
distinguishes them from classical ones is quantum entanglement.

Quantum entanglement is a fundamental property that is inherent in quantum systems \cite{EPRP,horodecki2009}. It is a key ingredient in the
implementation of various quantum information processes \cite{feynman1982,nielsen2000}. It opens new possibilities
which are not available for classical systems. The presence of quantum entanglement is a necessary component in physical
systems which are uses for realization of such schemes as quantum cryptography \cite{Ekert1991},
super-dense coding \cite{Bennett1992}, teleportation \cite{TELEPORT,Zeilinger1997}, optimization of quantum
calculations \cite{Giovannetti20031,Giovannetti20032,Batle2005,Borras2006}, etc. The implementation of these schemes
began after paper \cite{ASPECT}, where testing Bell's inequality \cite{BELL} Aspect et al. experimentally solved the
EPR paradox \cite{EPRP}. For instance, the simplest scheme of quantum information, where the entanglement plays an important role,
is a quantum teleportation of qubit state \cite{TELEPORT}. This scheme requires preparation of a
two-qubit entangled state. Moreover, to ensure full teleportation, this state should be maximally entangled.
However, the influence of the environment and device errors do not often allow to achieve the states with maximal
entanglement. This leads to incomplete teleportation of a qubit. Thus, one of the necessary step in the implementation of schemes
of quantum information is determination of the value of entanglement of the quantum system. It is important to develop protocols which allow us
to define the value of entanglement of quantum states.

Recently, a protocol which allows to determine the negativity as a measure of entanglement between qubits was
considered \cite{wang2018,mooney2019}. The authors studied entanglement of the states prepared on the 16-qubit \cite{wang2018} and 20-qubit \cite{mooney2019}
IBM Q quantum processors. They obtained that the state is inseparable with respect to any fixed pair
of qubits. In other papers the methods for determining the entanglement of states prepared on the 20-qubit ion trap system \cite{monz2011,friis2018},
system of photons \cite{wang2016,wang20181,zhong2018} and superconducting system \cite{song2017,gong2019} were considered.
Based on paper \cite{frydryszak2017}, we developed the protocol for determining the value of entanglement of a certain qubit
with the remaining system on a quantum computers \cite{kuzmak2020}. We applied this protocol for determining entanglement of pure states
prepared on IBM quantum devices. This protocol has a significant benefit for determining the entanglement of pure states,
because to determine entanglement of the system, the mean value of only one spin
should be measured. Due to the fact that only one spin should be measured, the readout errors in determining
the entanglement of the system is minimized. In addition, the entanglement in the special case of two-qubit rank-2 mixed state was studied.
The protocol \cite{kuzmak2020} was used for detemination the values of entanglement of different pure quantum states \cite{kuzmak20202,gnatenko2021}.

In this paper we study the entanglement of general rank-2 mixed quantum states prepared on a quantum
computer. We propose the protocol which allows one to determine the entanglement of a certain qubit
with the rest of the system in a rank-2 mixed state (Sec.~\ref{sec2}). This protocol is based on probing of the correlations
between qubits of the system. In addition, we obtain the expression which connects the concurrence of a two-qubit rank-2 mixed state
with these correlations (Sec.~\ref{sec3}). This is important because it allows us to measure the concurrence of such states
on quantum computers, which has not been done before. Using the protocol, in Sec.~\ref{sec4} we explore the entanglement of different
rank-2 mixed states prepared on the ibmq-melbourne quantum computer. Conclusions are given in Sec.~\ref{sec5}.

\section{Protocol for measurement of entanglement of a rank-2 mixed state on a quantum computer \label{sec2}}

Based on the definition of the geometry measure of entanglement of rank-2 mixed states \cite{frydryszak2017}, we develop the protocol
for measuring the value of entanglement between a certain qubit and the remaning system. This protocol requires the measurement
of the correlations between spins. Let us describe this protocol in detail. Thus, the rank-2 mixed state
is defined by the density matrix
\begin{eqnarray}
\rho=\sum_{\alpha} \omega_{\alpha}\vert\psi_{\alpha}\rangle\langle\psi_{\alpha}\vert,
\label{rank2densitymatrix}
\end{eqnarray}
where $\vert\psi_{\alpha}\rangle$ is pure states
spanned by vectors $\vert{\bf 0}\rangle=\vert 00\ldots 0\rangle$, $\vert{\bf 1}\rangle=\vert 11\ldots 1\rangle$
and $\sum_{\alpha} \omega_{\alpha}=1$. Then, in the case of an $N$-qubit state the value of entanglement
of the $i$th  qubit with the remaining system is determined by the expression obtained in paper \cite{frydryszak2017}
\begin{eqnarray}
E\left(\rho\right)=\frac{1}{2}\left(1-\sqrt{1-\langle\Sigma_i^x\rangle^2-\langle\Sigma_i^y\rangle^2}\right).
\label{mixedstateent}
\end{eqnarray}
Here, operators $\Sigma_i^x=\sigma_1^x\sigma_2^x\ldots\sigma_i^x\ldots\sigma_N^x$,
$\Sigma_i^y=\sigma_1^x\sigma_2^x\ldots\sigma_i^y\ldots\sigma_N^x$, $\Sigma_i^z=I_1I_2\ldots\sigma_i^z\ldots I_N$
play the role of the Pauli operators, which act on the subspace spanned by vectors
$\vert{\bf 0}\rangle$, $\vert{\bf 1}\rangle$, where $I_i$ is a unity single-qubit operator and $\sigma_i^{a}$ is
the $a$-component of the Pauli operator dependent on the qubit of number $i$.

The mean values in equation (\ref{mixedstateent}) can be expressed by the correlation functions as follows
\begin{eqnarray}
\langle\Sigma_i^a\rangle = \sum_{\alpha} \omega_{\alpha}\langle\psi_{\alpha}\vert\Sigma_i^a\vert\psi_{\alpha}\rangle,
\label{corrfuncx}
\end{eqnarray}
where $a=x,y$. Quantum computers allow us to provide the measurements of qubit states on the eigenstates
$\vert 0\rangle$, $\vert 1\rangle$ of the $\sigma^z$ operator. Thus, to measure the mean values of $\Sigma_i^x$ and $\Sigma_i^y$ operators
we represent the $\sigma^x$ and $\sigma^y$ operators in the following way
\begin{eqnarray}
\sigma^x=e^{-i\frac{\pi}{4}\sigma^y}\sigma^ze^{i\frac{\pi}{4}\sigma^y},\quad \sigma^y=e^{i\frac{\pi}{4}\sigma^x}\sigma^ze^{-i\frac{\pi}{4}\sigma^x}.
\label{xycomponents}
\end{eqnarray}
Using these representations, the correlation functions, which define expression (\ref{corrfuncx}), can be expressed as follows
\begin{eqnarray}
\langle\psi_{\alpha}\vert\Sigma_i^x\vert\psi_{\alpha}\rangle =\langle\tilde{\psi}^{yy\ldots y\ldots y}\vert\sigma_1^z\sigma_2^z\ldots\sigma_i^z\ldots\sigma_N^z\vert\tilde{\psi}^{yy\ldots y\ldots y}\rangle =p^x_+-p^x_-,\nonumber\\
\langle\psi_{\alpha}\vert\Sigma_i^y\vert\psi_{\alpha}\rangle =\langle\tilde{\psi}^{yy\ldots x\ldots y}\vert\sigma_1^z\sigma_2^z\ldots\sigma_i^z\ldots\sigma_N^z\vert\tilde{\psi}^{yy\ldots x\ldots y}\rangle =p^y_+-p^y_-,
\label{mvcorrfunc}
\end{eqnarray}
where $\vert\tilde{\psi}^{yy\ldots y\ldots y}\rangle=e^{i\frac{\pi}{4}\left(\sigma_1^y+\sigma_2^y\ldots +\sigma_i^y\ldots\sigma_N^y\right)}\vert\psi\rangle$,
$\vert\tilde{\psi}^{yy\ldots x\ldots y}\rangle=e^{i\frac{\pi}{4}\left(\sigma_1^y+\sigma_2^y\ldots -\sigma_i^x\ldots\sigma_N^y\right)}\vert\psi\rangle$.
From expressions (\ref{mvcorrfunc}) it follows that the qubit states should be rotated around the $x$- or $y$-axis by angles $\pi/2$.
To calculate mean values in formula (\ref{mvcorrfunc}) a sufficient large number of individual measurements should be made. The result of an individual measurement is the value $+1$ or $-1$.
Making a certain number of shots on quantum computer, we obtain that the part of measurements $p_+$ takes the value $+1$, and other part $p_-$ takes the value $-1$.

\begin{figure}[!!h]
\includegraphics[scale=0.650, angle=0.0, clip]{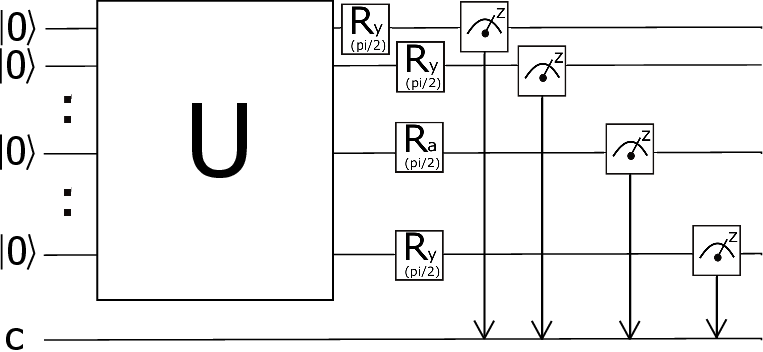}
\caption{Method for determining the correlations between qubits in the state generated from
the initial state $\vert \bf{0}\rangle$ to state $\vert\psi_{\alpha}\rangle$
by the unitary operator $U$. The $R_a$ is a gate which provides the rotation of the qubit around the predefined axise $a=x,y$ by the angle $\pi/2$.}
\label{mixingprot}
\end{figure}

Thus, to define the value of entanglement of a rank-2 mixed state consisting of $\vert\psi_{\alpha}\rangle$ pure states,
the correlation functions (\ref{mvcorrfunc}) of each pure state $\vert\psi_{\alpha}\rangle$ should be measured in the way shown in Fig.~\ref{mixingprot}.
Then, to obtain the mean values of operators $\Sigma_i^{\alpha}$, we substitute the correlations with the appropriate weights in expression (\ref{corrfuncx}).
Finally, using these mean values in definition (\ref{mixedstateent}), we obtain the value of entanglement
between a certain qubit and the remaining system.

\section{Measuring the concurrence by correlations \label{sec3}}

In paper \cite{Wootters1998} Wootters presented an explicit formula for finding the value of entanglement of a two-qubit state.
It is called concurrence and is denoted by $C$. In this section we express concurrence of a rank-2 mixed state
by the correlation functions considered in the previous section. This fact allows us to experimentally measure it values
for a predefined state.

The concurrence as a quantitative measure of entanglement of a particular two-qubit state
defined by density matrix $\rho$ can be calculated as follows
\begin{eqnarray}
C(\rho)=\max\{0,\lambda_1-\lambda_2-\lambda_3-\lambda_4\}.
\label{wootters}
\end{eqnarray}
Here, $\lambda_i$ are the eigenvalues, in decreasing order, of the Hermitian matrix $R=\sqrt{\sqrt{\rho}\tilde{\rho}\sqrt{\rho}}$, where
$\tilde{\rho}=\left(\sigma_1^y\sigma_2^y\right)\rho^*\left(\sigma_1^y\sigma_2^y\right)$. Note that $\lambda_i$ are real and positive numbers.
For calculations it is convenient to use the eigenvalues of the non-Hermitian matrix $\rho\tilde{\rho}$ which have the form $\lambda_i^2$.
In general case, it is impossible to find the protocol which would measure concurrence of any two-qubit mixed state
by correlation functions on a quantum computers. However, it is possible to find such a protocol for measuring
concurrence of a rank-2 mixed state of two qubits. Indeed, let us calculate the concurrence of state
(\ref{rank2densitymatrix}) with $\vert\psi_{\alpha}\rangle=a_{\alpha}\vert 00\rangle+b_{\alpha}\vert 11\rangle$.
Using definition (\ref{wootters}) for these state we obtain
\begin{eqnarray}
C(\rho)=2\left\vert\sum_{\alpha}\omega_{\alpha}a_{\alpha}b^*_{\alpha}\right\vert.
\label{woottersrank2state}
\end{eqnarray}
The detailed derivation of these expression is presented in Appendix \ref{appa}.
On the other hand, it is easy to verify that the mean values of $\Sigma_1^x=\sigma_1^x\sigma_2^x$, $\Sigma_1^y=\sigma_1^y\sigma_2^x$
in two-qubit state (\ref{rank2densitymatrix}) have the form
\begin{eqnarray}
\langle\Sigma_1^x\rangle=2\Re\left(\sum_{\alpha}\omega_{\alpha}a_{\alpha}b^*_{\alpha}\right),\quad \langle\Sigma_1^y\rangle=-2\Im\left(\sum_{\alpha}\omega_{\alpha}a_{\alpha}b^*_{\alpha}\right).
\label{rank2meanvalues}
\end{eqnarray}
Comparing expressions (\ref{woottersrank2state}) and (\ref{rank2meanvalues}) we obtain the expression for concurrence
of a two-qubit rank-2 mixed state by mean correlations. This expression has the form
\begin{eqnarray}
C(\rho)=\sqrt{\langle\Sigma_1^x\rangle^2+\langle\Sigma_1^y\rangle^2}.
\label{woottersrank2statemeanv}
\end{eqnarray}
Note that instead $\langle\sigma_1^y\sigma_2^x\rangle$, we can consider the $\langle\sigma_1^x\sigma_2^y\rangle$
in this expression. This is because the first mean value is used to calculate the entanglement of the first qubit
with the second one, and vice versa, the second mean value is used to calculate the entanglement of the second qubit
with the first one. However, this two cases describe the same value of entanglement between qubits.
In these two cases the values of concurrence are the same. Therefore, to determine the entanglement
between the two spins, we can measure mean value $\langle\sigma_1^y\sigma_2^x\rangle$ or $\langle\sigma_1^x\sigma_2^y\rangle$.
In addition, it should be noted that expression (\ref{woottersrank2statemeanv}) can be used
for any two-qubit rank-2 mixed state decomposed by another basis.
However, in this case the $\Sigma_1^x$, $\Sigma_1^y$ operators, which provide the transformation of the basis states
as the Pauli operators should be found. Then, these analogues of the Pauli operators can be substituted into
equation (\ref{woottersrank2statemeanv}).

Finally, as can be seen from equations
(\ref{mixedstateent}) and (\ref{woottersrank2statemeanv}), the concurrence and geometry measure of entanglement
of a two-qubit rank-2 mixed state are related by the expression
\begin{eqnarray}
C(\rho)=2\sqrt{E(\rho)\left(1-E(\rho)\right)}.
\label{geommeasrelwootters}
\end{eqnarray}
Thus, using the protocol for measuring the correlations between qubits, which is described in the previous section,
the concurrence of a two-qubit rank-2 mixed state can be measured on a quantum computer.

\section{Determining the entanglement of a rank-2 mixed states prepared on the ibmq-melbourne quantum computer \label{sec4}}

\begin{figure}[!!h]
\includegraphics[scale=0.700, angle=0.0, clip]{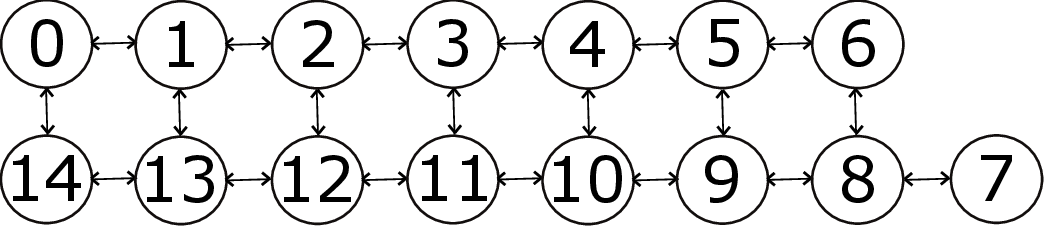}
\caption{The ibmq-melbourne quantum device consists of fifteen superconducting qubits which
interact between themselves in the way shown in the figure. The bidirectional arrows mean that the certain pair of qubits can be connected by the controlled-NOT operator
in a way that each of the qubit can be both a control and a target.}
\label{ibmq_16_melbourne}
\end{figure}

The IBM has developed a cloud service called the IBM Q Experience \cite{IBMQExp,OpenQasm},
which allows free access to different quantum devices operating on superconducting qubits. In this section
we study the entanglement of rank-2 mixed states prepared on the ibmq-melbourne quantum computer. This computer consists of fifteen superconducting
qubits which interact between themselves in the way shown in Fig.~\ref{ibmq_16_melbourne}. The algorithms prepared on this computer are
performed by four single-qubit and one two-qubit basis gates. These gates are
as follows: $I$, $U_1(\lambda)$, $U_2(\phi,\lambda)$, $U_3(\theta,\phi,\lambda)$,
and controlled-NOT gate (${\rm CX}$) \cite{OpenQasm}, where $\lambda\in[0,2\pi]$, $\theta\in[0,\pi]$ and $\phi\in[0,2\pi]$
are some real parameters. Using these basis gates, an arbitrary operation $U$ can be implemented. Let us describe these
gates in detail. The $I$ gate is a single-qubit unit operator. The $U_3(\theta,\phi,\lambda)$ operator is the most general
single-qubit gate which has the form
\begin{eqnarray}
U_3(\theta,\phi,\lambda)=e^{-i\phi \sigma^z/2}e^{-i\theta \sigma^y/2}e^{-i\lambda \sigma^z/2}.
\label{U3gatedeff}
\end{eqnarray}
It transforms the state $\vert 0\rangle$ to an arbitrary one-qubit state
\begin{eqnarray}
U_3(\theta,\phi,\lambda)\vert 0\rangle =e^{-i\lambda/2}\left(\cos\frac{\theta}{2}\vert 0\rangle+\sin\frac{\theta}{2}e^{i\phi}\vert 1\rangle\right).
\label{arboneqstate}
\end{eqnarray}
The other two single-qubit gates are a specific cases of the $U_3(\theta,\phi,\lambda)$ operator. They can be defined in the following way
\begin{eqnarray}
&&U_1(\lambda)=\exp(-i\lambda \sigma^z/2)=U_3(0,0,\lambda),\nonumber\\
&&U_2(\phi,\lambda)=e^{-i\phi \sigma^z/2}e^{-i\pi \sigma^y/4}e^{-i\lambda \sigma^z/2}=U_3(\pi/2,\phi,\lambda).
\label{U1andU2gates}
\end{eqnarray}
As we can see, each of the above gates provides the rotation of the qubit state around a certain axis defined
by the $\theta$, $\phi$ and $\lambda$ parameters. For instance, the $U_1(\lambda)$ gate rotates the qubit state around
the $z$-axis by the angle $\lambda$. The last basic gate of the IBM's quantum devices is the controlled-NOT gate. It acts on a pair of qubits, with one acting as
'control' and the other as 'target'. It provides a $\sigma^x$ gate on the target qubit whenever the control qubit
is in state $\vert 1\rangle$. It is important to note that if the control qubit is in a superposition state,
this gate creates an entangled state.

The ibmq-melbourne allows one to perform single-qubit basis operations with fidelity $>99.6\%$ and almost all controlled-NOT
operations with fidelity $>95\%$ (See Table~\ref{taberrorsmix}). As we can see, almost every
qubit can be read with fidelity $>95\%$.

Using our protocol let us study the entanglement of different rank-2 mixed states prepared on the ibmq-melbourne quantum device.
Note that in the case of a mixed state to obtain the value of entanglement, all the qubits of the system should be measured.
Therefore, in the case of a mixed state, the error of the results are greater than in the case of pure state,
where only one qubit of the system is measured. The IBM Q devices still are not
accurate to determine the entanglement of a mixed state consisting of a large number of qubits.
In this section we prepare 4- and 2-qubit mixed states which consist of Schr\"odinger cat states
and measure their value of entanglement. For this purpose we use q[0]-q[3] qubits because they have the smallest
readout and single-qubit errors (see Table~\ref{taberrorsmix}). They also have long coherence times.
The errors which appear by performing the controlled-NOT operations between these qubits are quite small.
We consider the rank-2 mixed state defined by density matrix
\begin{eqnarray}
\rho_{cat}=\omega\vert\psi_{cat}^+\rangle\langle\psi_{cat}^+\vert+\left(1-\omega\right)\vert\psi_{cat}^-\rangle\langle\psi_{cat}^-\vert,
\label{mixedSchcatstate}
\end{eqnarray}
where $\vert\psi_{cat}^{\pm}\rangle=1/\sqrt{2}\left(\vert {\bf 0}\rangle\pm \vert {\bf 1}\rangle\right)$, $\omega\in[0,1]$.
The mean values in state (\ref{mixedSchcatstate}) according to equation (\ref{corrfuncx})
are the following
\begin{eqnarray}
\langle \Sigma_1^x\rangle=2\omega-1,\quad \langle \Sigma_1^y\rangle=0.
\label{meanvaluesmixstate}
\end{eqnarray}
Then, using these mean values in definition (\ref{mixedstateent}), we obtain the geometric measure of entanglement between any qubit
and the remaining system in state (\ref{mixedSchcatstate}) as a function of parameter $\omega$
\begin{eqnarray}
E\left(\rho_{cat}\right)=\frac{1}{2}\left(1-2\sqrt{\omega(1-\omega)}\right).
\label{entmixedSchcatstate}
\end{eqnarray}
In the case of two qubits, using mean values (\ref{meanvaluesmixstate}) in expression (\ref{woottersrank2statemeanv}) we obtain
the concurrence of state (\ref{mixedSchcatstate})
\begin{eqnarray}
C\left(\rho_{cat}\right)=\left\vert 2\omega-1\right\vert.
\label{entmixedSchcatstateconc}
\end{eqnarray}

\begin{table}[h]
\caption{Calibration parameters of ibmq-melbourne quantum device, archived 23 April 2020
from reference \cite{IBMQExp}.}

\begin{tabular}{ c c c c c c c c c c c c c c c c }
& {\bf Q0} & {\bf Q1} & {\bf Q2} & {\bf Q3}  \\
\\
{\bf T$_1$, $\mu s$} & 64.9 & 49.9 & 55.3 & 70.0 \\
  \\
{\bf T$_2$, $\mu s$} & 22.6 & 83.1 & 120.7 & 56.4 \\
  \\
  {\bf Gate Error ($10^{-3}$)} & 0.75 & 0.96 & 1.19 & 0.50  \\
  \\
  {\bf Readout Error ($10^{-2}$)} & 1.85 & 2.70 & 2.15 & 3.70  \\
  \\
\multirow{1}{*}{{\bf Multi-Qubit Gate (CX)}}& {\bf 0\_1} & {\bf 1\_2} & {\bf 2\_3} &  \\
\multirow{1}{*}{{\bf Error ($10^{-2}$)}} & 2.65 & 1.48 & 2.99 &  \\
\end{tabular}

\label{taberrorsmix}
\end{table}

\begin{figure}[!!h]
\subfigure[]{\label{mixed_cat1}}{\includegraphics[scale=0.70, angle=0.0, clip]{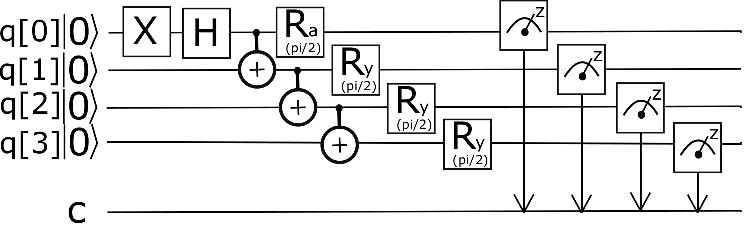}}
\subfigure[]{\label{mixed_cat2}}{\includegraphics[scale=0.70, angle=0.0, clip]{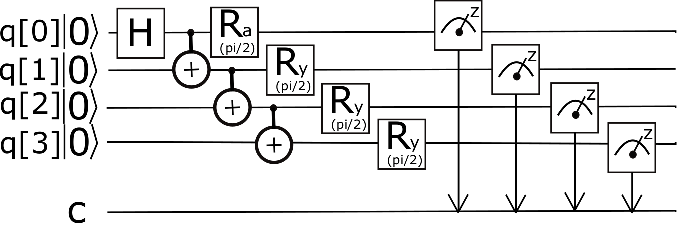}}
\caption{Quantum circuits for implementation and measurement of the correlations
of 4-qubit $\vert\psi_{cat}^-\rangle$ (a)
and $\vert\psi^+_{cat}\rangle$ (b) Schr\"odinger cat states.
The $R_a$ is a gate which provides the rotation of the qubit around the predefined axise $a=x,y$ by the angle $\pi/2$.}
\label{mixed_Bell}
\end{figure}

\begin{figure}[!!h]
\subfigure[]{\label{mixed_1qres}}{\includegraphics[scale=0.70, angle=0.0, clip]{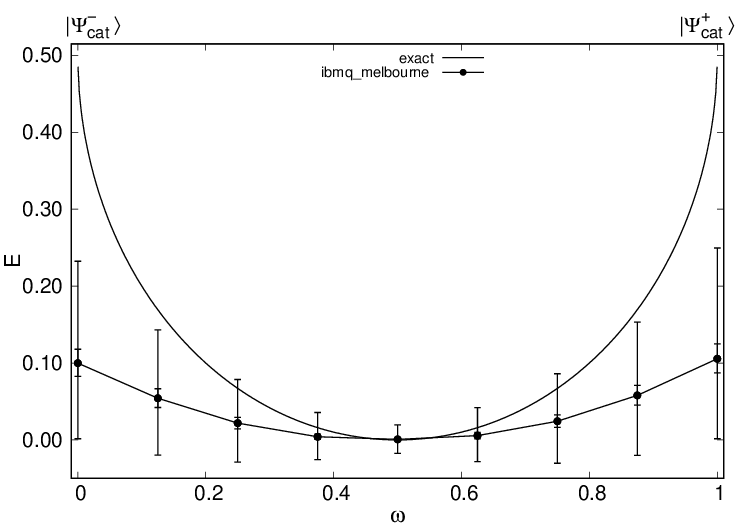}}
\subfigure[]{\label{mixed_2qres}}{\includegraphics[scale=0.70, angle=0.0, clip]{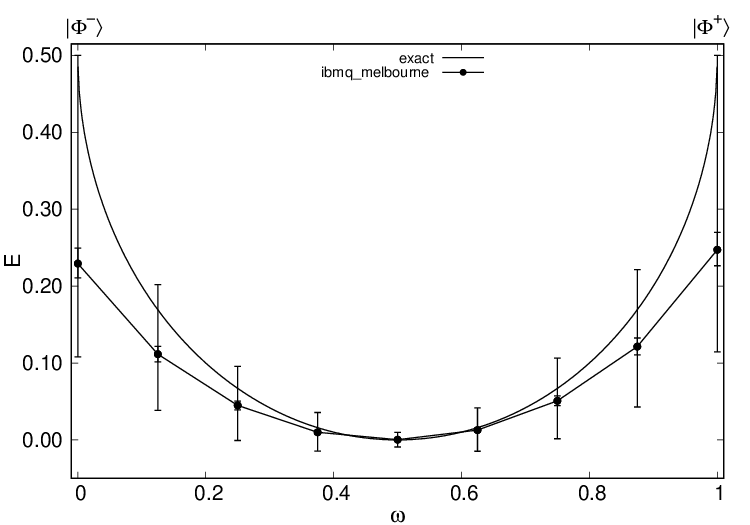}}
\caption{Dependence of the geometric measure of entanglement of mixed states which consist of 4-qubit
$\vert\psi^+_{cat}\rangle$, $\vert\psi^-_{cat}\rangle$ states (a),
2-qubit $\vert\Phi^+\rangle$, $\vert\Phi^-\rangle$ Bell states (b)
on the weight parameter $\omega$.}
\label{mixed_states_ent}
\end{figure}

\begin{figure}[!!h]
\includegraphics[scale=0.70, angle=0.0, clip]{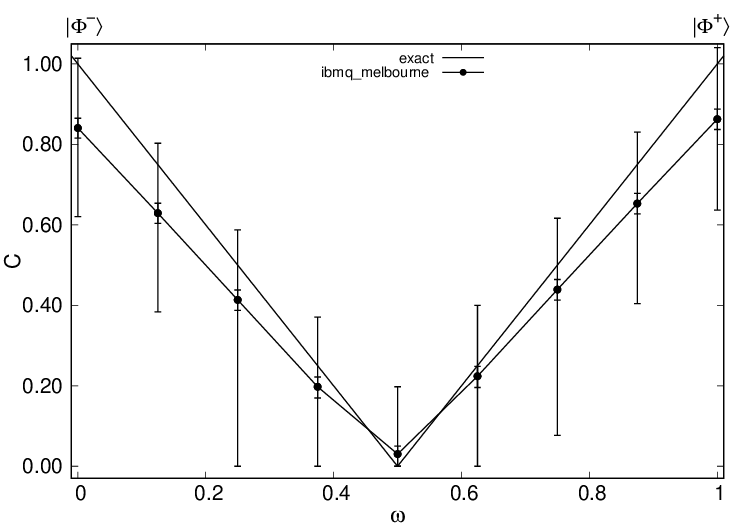}
\caption{Dependence of concurrence of a two-qubit mixed state, which consists of the Bell states
$\vert\Phi^+\rangle$, $\vert\Phi^-\rangle$, on the weight parameter $\omega$.}
\label{mixed_states_ent3}
\end{figure}

\begin{figure}[!!h]
\subfigure[]{\label{mixed_state_1}}{\includegraphics[scale=0.70, angle=0.0, clip]{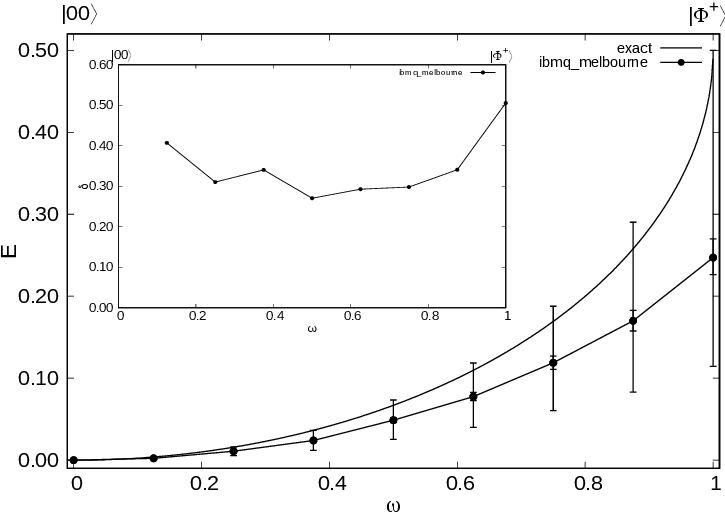}}
\subfigure[]{\label{mixed_state_2}}{\includegraphics[scale=0.70, angle=0.0, clip]{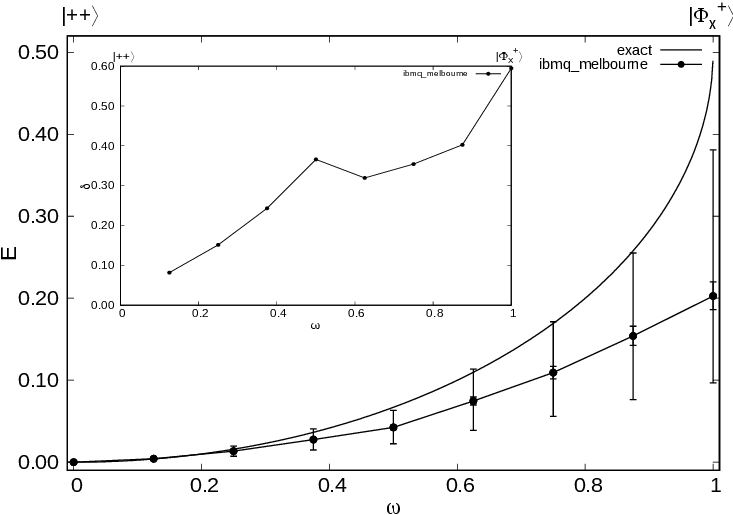}}
\caption{Dependence of the geometric measure of entanglement of mixed states which consist of
2-qubit  $\vert\Phi^+\rangle$, $\vert 00\rangle$ states (a) and the states, where $\vert\Phi^+\rangle$, $\vert 00\rangle$ are projected on the $x$-axes (b),
on the weight parameter $\omega$.
In the insets, the relative deviation between theoretical ($E$) and measured ($E_i$) values of entanglement,
$\delta=\vert E_i-E\vert/E$.}
\label{mixed_states_ent2}
\end{figure}

On the ibmq-melbourne quantum device we prepare and measure the 4-qubit
$\vert\psi^-_{cat}\rangle$ and $\vert\psi^+_{cat}\rangle$ (Fig.~\ref{mixed_Bell}) states.
The number of shots we set according to the weights defined by $\omega$.
We provide measurements of these states in the way such that the total number of shots remained equal to 8192.
Then, for predefined $\omega$ the numbers of measurements of each pure quantum states are defined in the way: $8192 \times\omega$ for $\vert\psi^+_{cat}\rangle$ state
and $8192 \times(1-\omega)$ for $\vert\psi^-_{cat}\rangle$. In our experiment, we change $\omega$ with step 0.125, which means that the number
of shots is changed with step 1024. According to this procedure, we study the entanglement of
a rank-2 mixed state (\ref{mixedSchcatstate}). The exact (\ref{entmixedSchcatstate}) and experimental dependences of geometric measure of entanglement
on parameter $\omega$ in the case of 4- and 2-qubit mixed states are presented in Fig.~\ref{mixed_states_ent}.
In addition, in Fig~\ref{mixed_states_ent3} we present the behavior of concurrence (\ref{entmixedSchcatstateconc})
of two qubit state (\ref{mixedSchcatstate}). The wide error bars show general errors which appear during the measurement of entanglement on the ibmq-melbourne quantum computer,
and the narrow error bars demonstrate the gate errors. We describe the error estimation in Appendix \ref{appb}. In Table~\ref{tabelerrors} we present the errors which appear
in the ibmq-melbourne quantum computer for different quantum states. As we can see, the main contribution in the discrepancies between the measured and theoretical results appears due to readout errors.
The readout error increases rapidly with the number of measured qubits. The more qubits are measured the faster the total readout error grows. Due to readout errors, the results
obtained in the 4-qubit case are worse than in the case of the 2-qubit state. The readout error in the case of entangled states is greater then in the case of separated states.
We also observe that for more separated states, the entanglement is determined more precisely. This can be explained by the fact that
the basis vectors $\vert {\bf 0}\rangle$, $\vert {\bf 1}\rangle$ are both the eigenstates of the density matrix
$1/2(\vert {\bf 0}\rangle\langle {\bf 0}\vert+\vert {\bf 1}\rangle\langle {\bf 1}\vert)$, which defines maximally mixed state (\ref{mixedSchcatstate})
(with $\omega=1/2$), and the basis states on which the measurements are performed.
Thus, the quantum computer provides the more accurate measurements, the closer the state of the system is to the states $\vert {\bf 0}\rangle$, $\vert {\bf 1}\rangle$.
We check this point for two different 2-qubit states. Namely, we measure the entanglement of the mixed state
$\rho_{1}=\omega \vert\Phi^+\rangle\langle\Phi^+\vert+(1-\omega)\vert 00\rangle\langle 00\vert$
(Fig.~\ref{mixed_state_1}) and the similar state
$\rho_{2}=\omega \vert\Phi_x^+\rangle\langle\Phi_x^+\vert+(1-\omega)\vert ++\rangle\langle ++\vert$ (Fig.~\ref{mixed_state_2}),
where $\vert\Phi^+\rangle$, $\vert 00\rangle$ states are projected on the $x$-axes.
Here, $\vert\Phi^+_x\rangle=1/\sqrt{2}\left(\vert ++\rangle+\vert --\rangle\right)$ and $\vert \pm\rangle =1/\sqrt{2}\left(\vert 0\rangle\pm\vert 1\rangle\right)$.
To analyze the results in these cases, we calculate and show in the insets of Fig.~\ref{mixed_states_ent2}
the relative deviation between theoretical ($E$) and measured ($E_i$) values of entanglement, $\delta=\vert E_i-E\vert/E$.
As we can see in Fig.~\ref{mixed_state_2}, the closer the state of system is to the state $\vert ++\rangle$,
the more accurate it is measured. In the case shown in Fig.~\ref{mixed_state_1} we do not observe such behavior
for state $\vert 00\rangle$. This is because before measuring, each qubits is rotated around the $x$- or $y$-axis by angle $\pi/2$
(see Fig.\ref{mixingprot}). The rotations around the $y$-axis of qubits lead to the fact that the state $\vert ++\rangle$
is transformed in the basis state  $\vert 00\rangle$ of the quantum computer. Another situation is in the case of
the state $\vert 00\rangle$. After the rotations of qubits around the $x$- and $y$-axises
the state $\vert 00\rangle$ ceases to remain the basis state of the quantum computer. Note that in the case
of $\rho_{2}$ state the role of the Pauli operators plays $\Sigma_1^x=\sigma_1^z\sigma_2^z$, $\Sigma_1^y=\sigma_1^y\sigma_2^z$,
$\Sigma_1^z=\sigma_1^xI_2$. In both cases of $\rho_1$ and $\rho_2$ the exact equation of entanglement is as follows
$E=(1-\sqrt{1-\omega^2})/2$.

\section{Conclusions \label{sec5}}

We have proposed the protocol for determining the entanglement between a certain qubit and the remaining system
in rank-2 mixed states prepared on a quantum computer. This protocol is based on the measurement of correlations between
all qubits of the system. It allows one to measure the correlations in a system consisting
of any number of qubits. However, since the readout errors of each qubit of modern quantum computers are still quite high,
with increasing the qubits in the system the results become significantly worse. Therefore, we have tested our protocol
on the 4-qubit and 2-qubit rank-2 mixed states prepared on the ibmq-melbourne quantum computer. These mixed states consist of
2- and 4-qubit Schr\"odinger cat states. We study the dependence
of the geometric measure of entanglement on the weight parameter $\omega$.
We observe a good agreement between the experimental result and theoretical prediction
in the case of the separated mixed states. We have observed that the closer the state of the system is to the basis state of
the quantum computer, the more accurate it is measured, and vice versa.
In addition, we have derived the expression which connects the concurrence
of a two-qubit rank-2 mixed state with correlations between qubits. This connection allows us to measure
the concurrence of such states on a quantum computers, which has not been done before. Using this relation we have obtained the
concurrence for a 2-qubit mixed state consisting of the Bell states $\vert\Phi^+\rangle$, $\vert\Phi^-\rangle$. As a result
we obtain a good agreement of experimental results with theoretical predictions (Fig.~\ref{mixed_states_ent3}).

\section{Acknowledgements}
The authors thank  Yuri Krynytskyi and Drs. Andrij Rovenchak, Taras Verkholyak for useful comments.
This work was partly supported by Project 77/02.2020 (No.~0120U104801) from National Research Foundation of Ukraine.

\begin{appendices}
\section{Derivation of concurrence for a two-qubit rank-2 mixed state \label{appa}}
\setcounter{equation}{0}
\renewcommand{\theequation}{A\arabic{equation}}

In this appendix using Wootters definition (\ref{wootters}) we obtain the expression for concurrence of a two-qubit
rank-2 mixed state. The two-qubit rank-2 density matrix (\ref{rank2densitymatrix})
with $\vert\psi_{\alpha}\rangle=a_{\alpha}\vert 00\rangle+b_{\alpha}\vert 11\rangle$ and its $\tilde{\rho}$ matrix
which are spanned by $\vert 00\rangle$, $\vert 01\rangle$,
$\vert 10\rangle$, $\vert 11\rangle$ basis have the form
\begin{eqnarray}
\rho=\left( \begin{array}{ccccc}
\sum_{\alpha}\omega_{\alpha}\vert a_{\alpha}\vert^2 & 0 & 0  & \sum_{\alpha}\omega_{\alpha}a_{\alpha}b^*_{\alpha}\\
0 & 0 & 0 & 0 \\
0 & 0 & 0 & 0 \\
\sum_{\alpha}\omega_{\alpha}a^*_{\alpha}b_{\alpha} & 0 & 0 & \sum_{\alpha}\omega_{\alpha}\vert b_{\alpha}\vert^2
\end{array}\right),\quad
\tilde{\rho}=\left( \begin{array}{ccccc}
\sum_{\alpha}\omega_{\alpha}\vert b_{\alpha}\vert^2 & 0 & 0  & \sum_{\alpha}\omega_{\alpha}a_{\alpha}b^*_{\alpha}\\
0 & 0 & 0 & 0 \\
0 & 0 & 0 & 0 \\
\sum_{\alpha}\omega_{\alpha}a^*_{\alpha}b_{\alpha} & 0 & 0 & \sum_{\alpha}\omega_{\alpha}\vert a_{\alpha}\vert^2
\end{array}\right).
\label{denmatrixmatrix}
\end{eqnarray}
Then, the matrix $\rho\tilde{\rho}$ can be expressed as follows
\begin{eqnarray}
\rho\tilde{\rho}=\left( \begin{array}{ccccc}
\sum_{\alpha,\beta}\omega_{\alpha}\omega_{\beta}\left(\vert a_{\alpha}\vert^2\vert b_{\beta}\vert^2+a_{\alpha}a^*_{\beta}b^*_{\alpha}b_{\beta}\right) & 0 & 0 & \sum_{\alpha,\beta}\omega_{\alpha}\omega_{\beta}\left(\vert a_{\alpha}\vert^2 a_{\beta}b^*_{\beta}+a_{\alpha}b^*_{\alpha} \vert a_{\beta}\vert^2\right) \\
0 & 0 & 0 & 0 \\
0 & 0 & 0 & 0 \\
\sum_{\alpha,\beta}\omega_{\alpha}\omega_{\beta}\left(a^*_{\alpha} b_{\alpha} \vert b_{\beta}\vert^2+\vert b_{\alpha}\vert^2 a^*_{\beta}  b_{\beta}\right) & 0 & 0 & \sum_{\alpha,\beta}\omega_{\alpha}\omega_{\beta}\left(\vert b_{\alpha}\vert^2\vert a_{\beta}\vert^2+a^*_{\alpha}a_{\beta}b_{\alpha}b^*_{\beta}\right)
\end{array}\right).
\label{rhotilderho}
\end{eqnarray}
The eigenvalues of this matrix
\begin{eqnarray}
\lambda_{1,2}^2=\left(\sqrt{\sum_{\alpha,\beta}\omega_{\alpha}\omega_{\beta}\vert a_{\alpha}\vert^2\vert b_{\beta}\vert^2}\pm \left\vert \sum_{\alpha}\omega_{\alpha}a_{\alpha}b^*_{\alpha}\right\vert\right)^2\quad \lambda_{3,4}^2=0.
\label{lambdas}
\end{eqnarray}
Now, substituting $\lambda_i$ in formula (\ref{wootters}) we obtain expression (\ref{woottersrank2state}).

\section{Errors of calculations \label{appb}}
\setcounter{equation}{0}
\renewcommand{\theequation}{A\arabic{equation}}

Total error $\Delta$ of calculations on a quantum computer consists of a standard error $\Delta_s$, an error of gates $\Delta_g$ and a readout error $\Delta_r$. Standard error is inversely proportional to the square root
of number of the shots. In our case, the number of shots is equal
to 8192, which makes the standard error very small with respect to other errors.The upper limit of the gate error consists of the sum of the errors of each gate included by the quantum circle (see, for instance, \cite{kitaev2002}).
Using the data from Table~\ref{taberrorsmix},
we can easily estimate this error for different quantum circles. In the cases of the 4-qubit Schr\"odinger cat (Fig.~\ref{mixed_Bell}) and Bell states these errors are equal to 0.07535 and 0.0290, respectively.
Here, we use the fact that the sequential action of single-qubit $\sigma^x$ and Hadamard gates is implemented using one single-qubit basis gate on a quantum computer.
Note that in the case of states projected on the $x$-axis the errors of the single-qubit operartors which provide rotations of qubits should be added. For a 2-qubit state
these rotations are provided by two single-qubit gates. The error, which appears from these gates, is equal to 0.0017. Finally, let us evaluate the readout error.
The readout error is presented in Table~\ref{taberrorsmix}, where only one qubit of the circle is measured. In the case of several qubits, the readout errors of each qubit cannot be added.
The more qubits are measured, the faster the total readout error grows. The readout error also depends on the measuring state. We suggest estimating the readout error of our calculations by using the average value
$F=\left(F_0+F_1\right)/2$ of fidelities $F_0$ and $F_1$ of the basis states
$\vert {\bf 0}\rangle$ and $\vert {\bf 1}\rangle$, respectively. The fidelities of achieving these states on the quantum computer are degraded by the readout error. Thus, the readout error is defined as follows $\Delta_r=1-F$,
where $F=\left(\vert a_0\vert^2+\vert a_1\vert^2\right)/2$ is the average fidelity, and $\vert a_0\vert$, $\vert a_1\vert$ are separately measured amplitudes which correspond to states $\vert {\bf 0}\rangle$, $\vert {\bf 1}\rangle$,
respectively. Here, $F_0=\vert a_0\vert^2$ and $F_1=\vert a_1\vert^2$. Then, for the 4-qubit and 2-qubit we obtain the readout errors equal to 0.3332 and 0.1773, respectively. In Table~\ref{tabelerrors} we represent the errors which appear
on the ibmq-melbourne quantum computer for different quantum circuits.

\begin{table}[h]
\caption{Errors which appear on the ibmq-melbourne quantum computers.}
\begin{tabular}{ c c c c c c c c c c c c c c c c }
State & $\Delta_g$ & $\Delta_r$ & $\Delta$ \\
\\
{\bf $\vert\psi^+_{cat}\rangle$} & 0.0754  & 0.3332  & 0.4086 \\
  \\
{\bf $\vert\Phi^+\rangle$} & 0.0290  & 0.1773  & 0.2063  \\
  \\
{\bf $\vert\Phi^+_x\rangle$} & 0.0307  & 0.1773 & 0.2080 \\
  \\
{\bf $\vert 00\rangle$} & 0 & 0.0892 & 0.0892 \\
  \\
{\bf $\vert ++\rangle$} & 0.0017  & 0.1773 & 0.1790  \\

\end{tabular}
\label{tabelerrors}
\end{table}

Now, using errors presented in Table~\ref{tabelerrors}, the errors of the value of entanglement can be obtained. Thus, for mean values (\ref{mvcorrfunc}) which we directly measure on the quantum computer we obtain the following restrictions:
$\langle\psi_{\alpha}\vert\Sigma^a_i\vert\psi_{\alpha}\rangle=\langle\psi_{\alpha}\vert\Sigma^a_i\vert\psi_{\alpha}\rangle\pm \langle\psi_{\alpha}\vert\Sigma^a_i\vert\psi_{\alpha}\rangle\Delta^{(\alpha)}$. Then, if we have two pure states
$\vert\psi_1\rangle$, $\vert\psi_2\rangle$ which define the mixed state, we obtain the restrictions for the square of mean value (\ref{corrfuncx}) in the form
\begin{eqnarray}
&&\langle\Sigma_i^a\rangle^2 = \langle\Sigma_i^a\rangle^2\nonumber\\
&&\pm \left[2\vert\langle\Sigma_i^a\rangle\vert\left(\omega \vert\langle \psi_1\vert\Sigma^a_i\vert\psi_1\rangle\vert \Delta^{(1)}+(1-\omega) \vert\langle \psi_2\vert\Sigma^a_i\vert\psi_2\rangle\vert\Delta^{(2)}\right)\right. \nonumber\\
&&\left.+\left(\omega \vert\langle \psi_1\vert\Sigma^a_i\vert\psi_1\rangle\vert\Delta^{(1)}+ (1-\omega) \vert\langle \psi_2\vert\Sigma^a_i\vert\psi_2\rangle\vert\Delta^{(2)}\right)^2 \right].
\label{restrictioncorrfuncx}
\end{eqnarray}
Finally, substituting these restrictions into expressions (\ref{mixedstateent}) and (\ref{woottersrank2statemeanv}) we find the deviations for the values of entanglement caused by the errors which appear on the ibmq-melbourne quantum computer.

\end{appendices}


\begin{thebibliography}{99}
\bibitem{feynman1982} R. P. Feynman, Int. J. Theor. Phys. {\bf 21}, 467 (1982).
\bibitem{lloyd1996} S. Lloyd, Science {\bf 273}, 1073 (1996).
\bibitem{buluta2009} I. Buluta, F. Nori, Science {\bf 326}, 108 (2009).
\bibitem{preskill2012} J. Preskill, arXiv:1203.5813 (2012).
\bibitem{krishnan2019} A. Krishnan, M. Schmitt, R. Moessner, M. Heyl, Phys. Rev. A {\bf 100}, 022125 (2019).
\bibitem{martinez2016} E. A. Martinez et al., Nature {\bf 534}, 516 (2016).
\bibitem{dumitrescu2018}E. F. Dumitrescu et al., Phys. Rev. Lett. {\bf 120}, 210501 (2018).
\bibitem{sugisaki2019} Kenji Sugisaki, Shigeaki Nakazawa, Kazuo Toyota, Kazunobu Sato, Daisuke Shiomi, Takeji Takui, ACS Cent. Sci. {\bf 5}, 167 (2019).
\bibitem{loss1998} D. Loss, D. P. DiVincenzo, Phys. Rev. A \textbf{57}, 120 (1998).
\bibitem{kane1998} B. E. Kane, Nature \textbf{393}, 133 (1998).
\bibitem{molmer1999} K. Molmer, A. Sorensen, Phys. Rev. Lett. {\bf 82}, 1835 (1999).
\bibitem{porras2004} D. Porras, J. I. Cirac, Phys. Rev. Lett. {\bf 92}, 207901 (2004).
\bibitem{duan2003} L.-M. Duan, E. Demler, M. D. Lukin, Phys. Rev. Lett. {\bf 91}, 090402 (2003).
\bibitem{kuklov2003} A. B. Kuklov, B. V. Svistunov, Phys. Rev. Lett. {\bf 90}, 100401 (2003).
\bibitem{gross2017} Ch. Gross, I. Bloch, Science {\bf 357}, 995 (2017).
\bibitem{wei2005} L. F. Wei, Yu-xi Liu and Franco Nori, Phys. Rev. B \textbf{71}, 134506 (2005).
\bibitem{mooij1999} J. E. Mooij, T. P. Orlando, L. Levitov, Lin Tian, Caspar H. van der Wal  and Seth Lloyd, Science \textbf{285}, 1036 (1999).
\bibitem{makhlin2001} Y. Makhlin, G. Sch\"on, A. Shnirman, Rev. Mod. Phys. \textbf{73}, 357 (2001).
\bibitem{majer2007} J. Majer et al., Nature \textbf{449}, 443 (2007).
\bibitem{EPRP} A. Einstein, B. Podolsky, N. Rosen, Phys. Rev. {\bf 47}, 777 (1935).
\bibitem{horodecki2009} R. Horodecki, P. Horodecki, M. Horodecki, K. Horodecki, Rev. Mod. Phys. {\bf 81}, 865 (2009).
\bibitem{nielsen2000} M. A. Nielsen, I. L. Chuang, {\it Quantum Computation and Quantum Information} (Cambridge University Press, Cambridge, 2000).
\bibitem{Ekert1991} A. K. Ekert, Phys. Rev. Lett., {\bf 67}, 661 (1991).
\bibitem{Bennett1992} Ch. H. Bennett, S. J. Wiesner, Phys. Rev. Lett. {\bf 69}, 2881 (1992).
\bibitem{TELEPORT} C. H. Bennett, G. Brassard, C. Crepeau, R. Jozsa, A. Peres, W. K. Wootters, Phys. Rev. Lett. {\bf 70}, 1895 (1993).
\bibitem{Zeilinger1997} D. Bouwmeester, J.-W. Pan, K. Mattle, M. Eibl, H. Weinfurter, A. Zeilinger, Nature {\bf 390}, 575 (1997).
\bibitem{Giovannetti20031} V. Giovannetti, S. Lloyd and L. Maccone, Europhys. Lett. {\bf 62}, 615 (2003).
\bibitem{Giovannetti20032} V. Giovannetti, S. Lloyd and L. Maccone, Phys. Rev. A {\bf 67}, 052109 (2003).
\bibitem{Batle2005} J. Batle, M. Casas, A. Plastino and A. R. Plastino, Phys. Rev. A {\bf 72}, 032337 (2005).
\bibitem{Borras2006} A. Borras, M. Casas, A. R. Plastino and A. Plastino, Phys. Rev. A {\bf 74}, 022326 (2006).
\bibitem{ASPECT} A. Aspect, J. Dalibard, G. Roger, Phys. Rev. Lett. {\bf 49}, 1804 (1982).
\bibitem{BELL} J. S. Bell, Physics {\bf 1}, 195 (1964).
\bibitem{wang2018} Yuanhao Wang, Ying Li, Zhang-qi Yin, Bei Zeng, npj Quant. Inf. {\bf 4}, 46 (2018).
\bibitem{mooney2019} G. J. Mooney, Ch. D. Hill, L. C. L. Hollenberg, Sci. Rep. {\bf 9}, 13465 (2019).
\bibitem{monz2011} T. Monz, et al., Phys. Rev. Lett. {\bf 106}, 130506 (2011).
\bibitem{friis2018} N. Friis, et al., Phys. Rev. X {\bf 8}, 021012 (2018).
\bibitem{wang2016} Xi-Li Wang, et al., Phys. Rev. Lett. {\bf 117}, 210502 (2016).
\bibitem{wang20181} Xi-Li Wang, et al., Phys. Rev. Lett. {\bf 120}, 260502 (2018).
\bibitem{zhong2018} Han-Sen Zhong, et al., Phys. Rev. Lett. {\bf 121}, 250505 (2018).
\bibitem{song2017} Chao Song, et al., Phys. Rev. Lett. {\bf 119}, 180511 (2017).
\bibitem{gong2019} Ming Gong, et al., Phys. Rev. Lett. {\bf 122}, 110501 (2019).
\bibitem{frydryszak2017} A. M. Frydryszak, M. I. Samar, V. M. Tkachuk, Eur. Phys. J. D {\bf 71}, 233 (2017).
\bibitem{kuzmak2020} A. R. Kuzmak, V. M. Tkachuk, Phys. Lett. A {\bf 384}, 126579 (2020).
\bibitem{kuzmak20202} A. R. Kuzmak, V. M. Tkachuk, Condens. Matter Phys. {\bf 23}, 43001 (2020).
\bibitem{gnatenko2021} Kh. P. Gnatenko, V. M. Tkachuk, Phys. Lett. A {\bf 396}, 127248 (2021).
\bibitem{IBMQExp} IBM Q Experience. https://quantum-computing.ibm.com.
\bibitem{OpenQasm} A. W. Cross, L. S. Bishop, J. A. Smolin, J. M. Gambetta, arXiv: 1707.03429 (2017).
\bibitem{Wootters1998} W. K. Wootters, Phys. Rev. Lett. {\bf 80}, 2245 (1998).
\bibitem{kitaev2002} A. Yu. Kitaev, A. H. Shen, M. N. Vyalyi, {\it Classical and Quantum Computation} (American Mathematical Society Providence, Rhode Island, 2002).
\end{thebibliography}
\end{document}